\documentclass[twocolumn]{aastex701}
\usepackage{amsmath}
\usepackage[version=4]{mhchem}
\graphicspath{{./}{newfig/}}

\newcommand{\xsol}{\,\times\,\mathrm{solar}}
\newcommand{\nel}{n_{\mathrm e}}
\newcommand{\nak}{\mathrm{(Na\!+\!K)/Cl}}

\begin{document}

\shorttitle{Jupiter's Deep Alkali--Chlorine Relation}
\title{Juno Microwave Observations Reveal Jupiter's Deep Alkali--Chlorine Relation}

\author[orcid=0000-0002-8706-6963]{Xi Zhang}
\affiliation{Department of Earth and Planetary Sciences, University of California, Santa Cruz, CA 95064, USA}
\email[show]{xiz@ucsc.edu}

\author[orcid=0000-0002-6666-5457]{Jiheng Hu}
\affiliation{Department of Climate and Space Sciences and Engineering, University of Michigan, Ann Arbor, MI 48109, USA}
\email{jihenghu@umich.edu}

\author[orcid=0000-0002-8280-3119]{Cheng Li}
\affiliation{Department of Climate and Space Sciences and Engineering, University of Michigan, Ann Arbor, MI 48109, USA}
\email{chengli@umich.edu}

\author[orcid=0009-0005-3389-8819]{Louis Siebenaler}
\affiliation{Leiden Observatory, University of Leiden, Einsteinweg 55, 2333CA Leiden, The Netherlands}
\email{siebenaler@strw.leidenuniv.nl}

\author[orcid=0000-0002-5104-8077]{Yury Aglyamov}
\affiliation{Department of Climate and Space Sciences and Engineering, University of Michigan, Ann Arbor, MI 48109, USA}
\email{ayury@umich.edu}

\begin{abstract}
The longest-wavelength channel of the \textit{Juno} Microwave Radiometer (MWR) probes Jupiter's kilobar
atmosphere through free electrons produced by sodium and potassium ionization. Under equilibrium chemistry the
electron abundance is the small residual of the charge balance between alkali cations and the anions \ce{Cl-} and
\ce{HS-}. Chlorine is not directly measurable in Jupiter's deep atmosphere because gaseous \ce{HCl} is removed from
the observable atmosphere by \ce{NH4Cl} condensation, whereas sulfur has been measured by the \textit{Galileo}
probe. The MWR-derived electron measurement therefore constrains the alkali-to-chlorine ratio rather than the alkali
abundance alone. We combine the MWR observations with equilibrium chemistry and microwave radiative transfer in a
Bayesian framework, finding that the deep gas-phase elemental alkali-to-chlorine abundance ratio is
$\nak\approx0.05$ over $0.3$--$5\xsol$ in chlorine, about $180$ times below the protosolar ratio of $8.7$. At
$3\xsol$ chlorine, the inferred alkali metallicity is $1.6\times10^{-2}\xsol$
($1\sigma$: $1.2\times10^{-2}$--$2.7\times10^{-2}\xsol$), while at low chlorine abundance \ce{HS-} sets an alkali
floor near $10^{-3}\xsol$. The inferred gas-phase alkali abundance exceeds the $\sim10^{-5}\xsol$ threshold by more than two orders of magnitude and rules out the long-proposed global kilobar radiative zone. Because sodium and potassium are refractory whereas chlorine is volatile, the inferred ratio provides a new diagnostic of the rock-to-ice balance in the solids accreted by Jupiter. This compositional interpretation assumes equilibrium chemistry; if lofted mineral clouds instead control the electron abundance under disequilibrium conditions, the inferred alkali--chlorine relationship need not hold.

\end{abstract}


\section{Introduction}\label{sec:intro}

Elemental abundances in Jupiter provide key constraints on its formation, yet its elemental inventory remains
incomplete. The \textit{Galileo} probe measured carbon, nitrogen, sulfur, and the noble gases in Jupiter's
atmosphere, finding enrichments of approximately three times protosolar
\citep{wongUpdatedGalileoProbe2004,atreyaDeepAtmosphereComposition2020}. Two important elements in the deep
atmosphere that remain absent from this inventory are the alkali metals (sodium and potassium) and chlorine. Their elemental
abundances would constrain the relative contributions of rocky and icy material to the solids that enriched Jupiter.

Sodium and potassium are difficult to measure by conventional remote sensing because they form salt clouds below the levels accessible to spectroscopy, including NaCl near 400~bar and KCl near 700~bar \citep{bhattacharyaHighlyDepletedAlkali2023}. Their deep gas-phase abundance also bears directly on an important question regarding Jupiter's interior structure. Jupiter's molecular envelope is usually assumed to remain convective to a great depth, but \citet{guillotNonadiabaticModelsJupiter1994} proposed that a stable radiative zone could form near kilobar pressures if the opacity becomes sufficiently low near $2000$~K. Subsequent opacity calculations showed that gaseous Na and K provide strong absorption at these temperatures, maintaining convection unless the deep alkali abundance is strongly depleted \citep{guillot2004interior,freedmanLineContinuumOpacities2008}. The most recent calculations show that a stable global radiative zone requires gas-phase alkalis below $\sim10^{-5}\xsol$ \citep{siebenalerAlkaliLinesExtreme2026}.

Chlorine is likewise difficult to measure because it is strongly depleted from the upper troposphere. Searches for gaseous \ce{HCl} have yielded upper mixing ratio limits of $\lesssim(2$--$5)\times10^{-9}$ from submillimeter and
thermal-infrared observations \citep{weissteinSubmillimeterLineSearch1996,fouchetUpperLimitsHydrogen2004}, and a
mole fraction below $6\times10^{-11}$ from \textit{Herschel}/PACS \citep{gappAbundancesTraceConstituents2024}. These
limits are far below the $\sim10^{-6}$ abundance expected if upper atmospheric chlorine were enriched similarly to Jupiter's other
measured volatiles. This non-detection is expected because \ce{HCl} reacts with abundant ammonia,
\[
\ce{NH3 + HCl -> NH4Cl(s)},
\]
and condenses near the 15--20~bar cloud base
\citep{weidenschillingAtmosphericCloudStructures1973,fegleyChemicalModelsDeep1994,showmanHydrogenHalidesJupiter2001}.
This sink acts faster than vertical transport can resupply \ce{HCl} from below
\citep{showmanHydrogenHalidesJupiter2001}. Chlorine is consequently sequestered in \ce{NH4Cl} as well as, deeper in the
atmosphere, in alkali chlorides, well below the observable troposphere.

The \textit{Juno} Microwave Radiometer (MWR; \citealt{janssenMWRMicrowaveRadiometer2017}) provides a window into
this otherwise inaccessible region. Its longest-wavelength channel (0.6~GHz; 50~cm) probes down to pressures of
1000--2000~bar, where free electrons from the thermal ionization of sodium and potassium contribute to the microwave
opacity \citep{bhattacharyaHighlyDepletedAlkali2023}. The observed electron abundance therefore provides an indirect
constraint on deep alkali chemistry. Under the assumption that alkalis are the only relevant electron donors and
sinks, a Saha calculation attributed the electron deficit to strongly subsolar alkali abundances in this 1-2 kilobar region,
$\sim10^{-2}$--$10^{-5}\xsol$ \citep{bhattacharyaHighlyDepletedAlkali2023}. In an equilibrium-rainout model that
includes the anions \ce{Cl-} and \ce{HS-}, \citet{aglyamovAlkaliMetalDepletion2025} instead inferred a local alkali
abundance of $\sim0.1\xsol$, assuming a chlorine abundance of $3\xsol$. Alternatively,
\citet{zhangAlkaliMetallicity2025} showed that mineral chemistry or dust--plasma interactions from vertically mixed clouds can
suppress the electron abundance even when the deep alkali abundance is solar or supersolar. This disequilibrium scenario, however, depends on highly uncertain microphysics and mineral chemistry.

In the equilibrium-rainout model, the electron abundance is controlled by the charge balance with \ce{Na+}, \ce{K+},
\ce{Cl-}, and \ce{HS-}. The alkali cations are balanced mainly by \ce{Cl-} and \ce{HS-}, leaving free
electrons as a small residual \citep{aglyamovAlkaliMetalDepletion2025}. Sulfur is constrained by the \textit{Galileo} measurement, whereas chlorine is not.
The MWR data therefore do not determine Jupiter's alkali abundance alone; rather, they constrain the combination of
alkali and chlorine abundances that reproduces the observed electron signal. This is not a direct detection of
chlorine, but it makes chlorine the key unknown in the equilibrium interpretation of the MWR data.

Here we combine \textit{Juno} brightness-temperature and limb-darkening measurements with equilibrium chemistry and
microwave radiative transfer to derive the joint alkali--chlorine constraint in a Bayesian framework. We compare our
equilibrium interpretation with previous studies and discuss its implications for Jupiter. Throughout this paper,
abundances are expressed relative to the protosolar values of \citet{asplundChemicalMakeupSun2021}, as compiled by
\citet{guillotGiantPlanetsInsideOut2023}; we refer to these values as ``solar.''\footnote{We
distinguish two notations. An \emph{abundance ratio} such as $\nak$ is the ratio of the elemental number
densities themselves. A \emph{metallicity} follows the astrophysical convention
$[\mathrm{M/H}]=\log_{10}(\mathrm{M/H})_{\rm planet}-\log_{10}(\mathrm{M/H})_\odot$, so that
$[\mathrm{M/H}]=0$ is solar and, for example, a $3\xsol$ enrichment is $[\mathrm{M/H}]=0.48$. Sodium and
potassium are varied together through a single enrichment factor, so
$[\mathrm{(Na\!+\!K)/H}]=[\mathrm{Na/H}]=[\mathrm{K/H}]$.}

\begin{figure*}[t]
\centering
\includegraphics[width=0.8\textwidth]{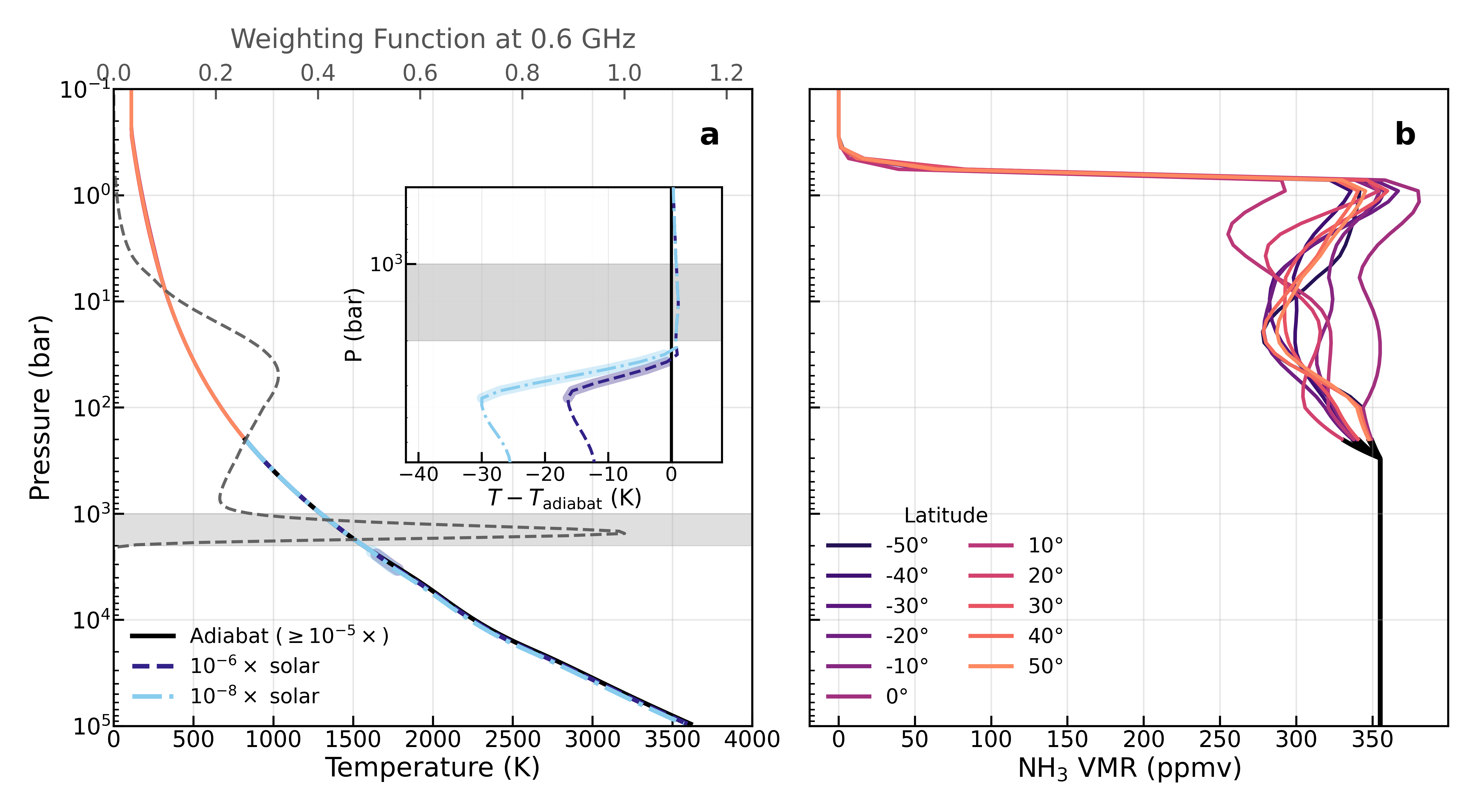}
\caption{Atmospheric structure adopted in this study. (a) Temperature profiles. Above 200~bar we
adopt the \textit{Juno} MWR retrieval of \citet{liSuperadiabaticTemperatureGradient2024}. Below
200~bar we use the deep temperature profiles of
\citet{siebenalerAlkaliLinesExtreme2026}: the adiabat for alkali $\ge10^{-5}\xsol$
and the subadiabatic profiles at $10^{-6}$ and $10^{-8}\xsol$,
where a radiative zone develops. The inset shows the temperature deviation from the adiabat,
$T-T_{\rm adiabat}$---the radiative zones lie at 2--4~kbar, below the 1000--2000~bar sensing window
(gray band), and are nearly indistinguishable from the adiabat where the MWR is sensitive.
The
dashed curve (top axis) shows the 0.6~GHz weighting function, with a shallow peak near
$\sim80$~bar dominated by \ce{NH3} and \ce{H2O}, and a deep peak near 1000--2000~bar dominated by
free electrons (gray band). Because this deep alkali-bearing layer contributes only
$\sim10$--20\% of the nadir emission, the limb darkening provides most of the constraint on the
deep electron abundance. (b) Retrieved \ce{NH3} volume mixing ratio as a function of pressure and
latitude from \citet{liSuperadiabaticTemperatureGradient2024}. Below $\sim200$~bar we assume
horizontally uniform deep abundances of 355~ppm for \ce{NH3} and 2573~ppm for \ce{H2O}.}
\label{fig:structure}
\end{figure*}

\section{Observations and Models} \label{sec:method}

\subsection{Juno Data}

We use the \textit{Juno} MWR dataset spanning 61 perijoves (PJ1--PJ61) and focus on the Channel~1
($0.6$~GHz) nadir brightness temperature $T_b(\phi)$ and limb darkening
$L_d(\phi)=[T_b(0^\circ)-T_b(45^\circ)]/T_b(0^\circ)\times100\%$ as functions of latitude $\phi$, binned
in $2^\circ$ intervals over $|\phi|\le60^\circ$. A preliminary reduction of the $T_b(\phi)$ data was
presented in \citet{zhangAlkaliMetallicity2025}. Following the approach of
\citet{oyafusoAngularDependenceSpatial2020}, \citet{zhangResidualStudyTesting2020}, and
\citet{liSuperadiabaticTemperatureGradient2024}, we fit a limb-darkening function to the deconvolved
footprint brightness temperatures at each perijove and latitude, reject non-thermal synchrotron and
lightning contamination through a multi-channel veto, apply the gravity correction, and combine the
surviving perijoves by their per-latitude median to obtain $T_b(\phi)$ and $L_d(\phi)$. Using the first
12 perijoves, \citet{liSuperadiabaticTemperatureGradient2024} retrieved the temperature and \ce{NH3}
distributions above 200~bar from the shallower MWR channels (2--6; Figure~\ref{fig:structure}); these
profiles constrain the weather-layer variability that shapes the latitudinal structure of our Channel~1
data.

The 0.6~GHz channel has a bimodal weighting function, with a shallow peak near 80~bar controlled by
\ce{NH3} and \ce{H2O}, and a deep peak near 1000--2000~bar controlled by free electrons
(Figure~\ref{fig:structure}a). The deep layer contributes only $\sim10$--20\% of the nadir brightness
temperature. The microwave opacity of \ce{NH3} under these deep conditions ($T\gtrsim500$~K) remains
poorly constrained, because laboratory measurements extend only to $\sim500$~K and
$\sim100$~bar \citep{hanleyNewModelHydrogen2009,bellottiLaboratoryMeasurements5202016}. As a result,
the nadir brightness temperature alone cannot uniquely determine the deep electron abundance.
Fortunately, the limb darkening is largely insensitive to the uncertain deep \ce{NH3} opacity
\citep{bhattacharyaHighlyDepletedAlkali2023}. Together, $T_b$ and $L_d$ provide a robust constraint on
the deep electron density.

\subsection{Chemistry and Radiative Transfer}

Our chemistry and radiative-transfer models follow
\citet{zhangAlkaliMetallicity2025}. For a given elemental composition, the equilibrium chemistry code
\texttt{GGchem} \citep{woitkeEquilibriumChemistry1002018} computes the abundances of all gas-phase
species, including the free electrons, assuming thermochemical equilibrium, rainout of condensed species, and an ideal gas mixture. This treatment neglects non-ideal fugacity corrections and pressure-induced modifications to the chemical network, such as level dissolution and pressure ionization. These effects could modify individual species abundances at kilobar pressures, but are not included in the present calculation. In this framework, deep mineral clouds remain sequestered below the observable atmosphere and
are not mixed upward into the MWR-sensitive region. The resulting electron profiles are passed to the
High-performance Atmospheric Radiation Package
(\texttt{HARP}; \citealt{liHighperformanceAtmosphericRadiation2018,bhattacharyaHARPRadiationPackage2023}),
which computes the microwave radiative transfer using the Appleton--Hartree cold-plasma refractive
index \citep{bhattacharyaHighlyDepletedAlkali2023}.

Above the 200~bar pressure level, we adopt the temperature and \ce{NH3} distributions retrieved from the MWR observations
(Figure~\ref{fig:structure}). Below 200~bar, the atmosphere is assumed to be horizontally homogeneous,
with fixed deep abundances of \ce{NH3} (355~ppm) and \ce{H2O} (2573~ppm). We fix sulfur to the
\textit{Galileo} value ($\mathrm{H_2S}/\mathrm{H_2}=8.9\times10^{-5}$, $2.9\xsol$;
\citealt{wongUpdatedGalileoProbe2004}) and treat the deep chlorine abundance as a free parameter. The
alkali abundance is a single free parameter applied jointly to sodium and potassium. For each
alkali--chlorine pair, \texttt{GGchem} computes the equilibrium abundances of \ce{K+}, \ce{Na+}, \ce{Cl-},
\ce{HS-}, and free electrons. The calculation assumes rainout of condensates and excludes the vertical transport of
mineral clouds, dust-catalyzed recombination, and dust-plasma interaction considered by
\citet{zhangAlkaliMetallicity2025}.

The deep temperature profile below the 200~bar level also depends on the alkali abundance through the possible formation of a
radiative zone. We therefore incorporate the self-consistent temperature structures calculated by
\citet{siebenalerAlkaliLinesExtreme2026}. Above the radiative-zone threshold ($10^{-5}\xsol$ alkali), we adopt their non-ideal
adiabatic profile. At lower alkali abundances, we use their subadiabatic temperature profiles in which a radiative zone develops
(Figure~\ref{fig:structure}a). Our forward model therefore consistently accounts for the feedback
between alkali abundance and the deep thermal structure. In practice this feedback does not affect the
result---the posterior ridge lies two to three orders of magnitude above the $10^{-5}\xsol$ threshold (Section \ref{subsec:alkali-cl}), so
the inference never enters the subadiabatic regime.

\begin{figure*}[t]
\centering
\includegraphics[width=0.95\textwidth]{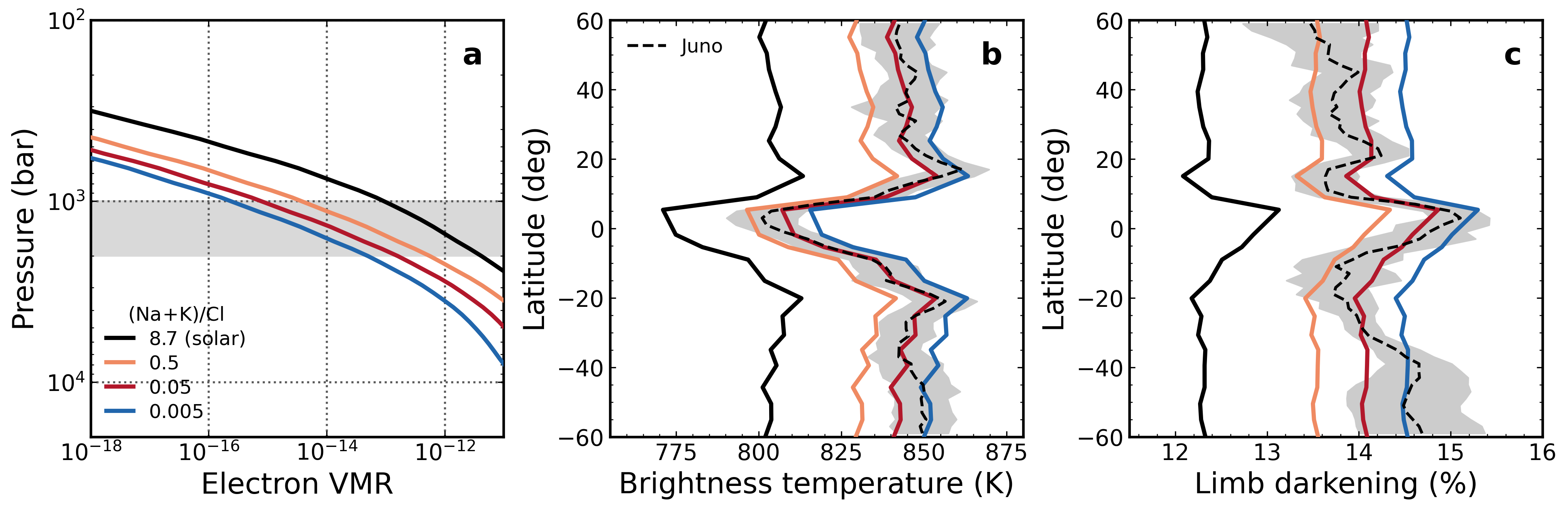}
\caption{Forward-model predictions compared with the \textit{Juno} MWR 0.6~GHz observations,
parameterized by the elemental alkali-to-chlorine abundance ratio $\nak$ at a fixed deep chlorine
abundance of $3\xsol$. Four ratios are shown: $8.7$ (the protosolar value, i.e.\ alkali and
chlorine equally enriched), $0.5$, $0.05$ (the best fit), and $0.005$. (a) Free-electron volume mixing ratio profiles; the gray band marks the
1000--2000~bar region to which the 0.6~GHz channel is most sensitive. (b) Nadir brightness temperature and
(c) limb darkening as functions of latitude; dashed curves and gray bands are the \textit{Juno} measurements
and their $1\sigma$ uncertainties. The equally enriched case ($\nak=8.7$) is strongly excluded. For the remaining ratios, the limb darkening provides the strongest discrimination and selects $\nak\approx0.05$, whereas the brightness temperature is additionally sensitive to the uncertain high-temperature \ce{NH3} opacity and absolute radiometric calibration. The modeled $T_b$ shown here includes the best-fit $+1.6\%$ coherent calibration offset.}
\label{fig:model}
\end{figure*}

\subsection{Bayesian Inference}
We introduce a Bayesian framework to constrain the deep electron density and the joint alkali--chlorine relation from
the \textit{Juno} MWR observations. For each model, we compute the brightness temperature $T_b(\phi)$ and limb darkening
$L_d(\phi)$ and compare them with the \textit{Juno} observations using a Gaussian likelihood
\[
\chi^2=
\sum_\phi
\left(
\frac{T_b^{\rm mod}-T_b^{\rm obs}}
{\sigma_{T_b}}
\right)^2
+
\sum_\phi
\left(
\frac{L_d^{\rm mod}-L_d^{\rm obs}}
{\sigma_{L_d}}
\right)^2,
\]
where the per-latitude uncertainties combine two independent contributions in quadrature. The first uncertainty is
the zonal scatter of the deconvolved 61-perijove ensemble, which measures the perijove-to-perijove
variability of the upper atmosphere. The second is the retrieval uncertainty of the deep temperature
and \ce{NH3} profiles, which we propagate through the forward model by sampling the retrieval posterior
of \citet{liSuperadiabaticTemperatureGradient2024}---the retrieved temperature and its coupled \ce{NH3}
opacity shift the modeled $T_b$ and $L_d$ together down the adiabat. The combined uncertainties are
$\sigma_{T_b}\approx8.5$~K and $\sigma_{L_d}\approx0.4\%$, of which the retrieval term contributes
$\approx7$~K and $\approx0.1\%$.

We account for two nuisance parameters: the high-temperature \ce{NH3} opacity scale $s$ (lognormal prior, $\sigma_s=0.3$~dex) and a coherent brightness-temperature calibration offset $\delta_{T_b}$ (Gaussian prior, $2\%$ of $T_b$), the latter matching the absolute radiometric calibration accuracy of the MWR \citep{janssenMWRMicrowaveRadiometer2017,liSuperadiabaticTemperatureGradient2024}. The opacity scale $s$ accounts for the uncertain \ce{NH3} microwave opacity above 500~K, where laboratory measurements are unavailable. It primarily affects the nadir brightness temperature, whereas the limb darkening is nearly insensitive to it. The offset $\delta_{T_b}$ scales the modeled brightness temperature coherently across all latitudes; the limb darkening, as a relative nadir-to-slant measurement, largely cancels this offset and therefore provides the more robust constraint \citep{liSuperadiabaticTemperatureGradient2024}. The deep-temperature uncertainty, previously imposed as a separate offset $\delta T$, enters through the retrieved temperature posterior folded into $\sigma_{T_b}$ and $\sigma_{L_d}$.

We evaluate the joint posterior on a two-dimensional grid of alkali and chlorine metallicities. At each grid point, the equilibrium-chemistry and radiative-transfer model predicts the MWR observables, and we marginalize over $s$ and $\delta_{T_b}$ by Gaussian-weighted integration on a fine grid. We adopt flat priors in $\log\mathrm{alkali}$ and $\log\mathrm{Cl}$. When quoting an alkali abundance at a specified chlorine metallicity, we condition this joint calculation on that chlorine value. Such conditional values, including the illustrative case of $3\xsol$ chlorine, are neither measurements of chlorine nor chlorine-marginalized alkali abundances.

\section{Results} \label{sec:results}

\subsection{The Deep Electron Density}

The \textit{Juno} MWR 0.6~GHz observations are sensitive to the deep electron density at pressures of 1000--2000~bar. For the illustrative model in Figure~\ref{fig:model}, we adopt a chlorine abundance of $3\xsol$, comparable to the measured enrichment of Jupiter's other volatiles, and vary the alkali-to-chlorine ratio. The electron abundance strongly affects the modeled averages of $T_b$ and $L_d$, whereas their latitudinal variations arise mainly from upper-atmosphere temperature and \ce{NH3} structure above 200~bar. The observations, with $L_d\approx14\%$, are reproduced by an electron number density of $\nel\sim1.4\times10^{14}~\mathrm{m^{-3}}$ at the weighting-function peak near 1500~bar, corresponding to a volume mixing ratio of $\sim2\times10^{-14}$.

\subsection{The Equilibrium Alkali--Chlorine Relation} \label{subsec:alkali-cl}

The electron density inferred above has a direct chemical implication in the equilibrium-rainout model. Near the 1500~bar weighting-function peak, \ce{K+} is the dominant cation and \ce{Cl-} is the dominant anion. The \ce{Na+} and \ce{K+} abundances track the total alkali abundance \citep{bhattacharyaHighlyDepletedAlkali2023}, whereas chlorine and sulfur capture electrons into \ce{Cl-} and \ce{HS-}, respectively \citep{aglyamovAlkaliMetalDepletion2025}. Charge neutrality therefore gives
\[
\nel \simeq n_{\mathrm{\ce{K+}}} + n_{\mathrm{\ce{Na+}}} - n_{\mathrm{\ce{Cl-}}} - n_{\mathrm{\ce{HS-}}} 
\]
where $n$ denotes the number density. The total cation and anion abundances agree to within about ten percent, leaving the free electrons as a small residual. At these pressures, \ce{Cl-} exceeds \ce{HS-} by a factor of $\sim\!1.5$; with sulfur fixed by the \textit{Galileo} measurement, chlorine is therefore the dominant unknown negative charge carrier.

\begin{figure}[t]
\centering
\includegraphics[width=\columnwidth]{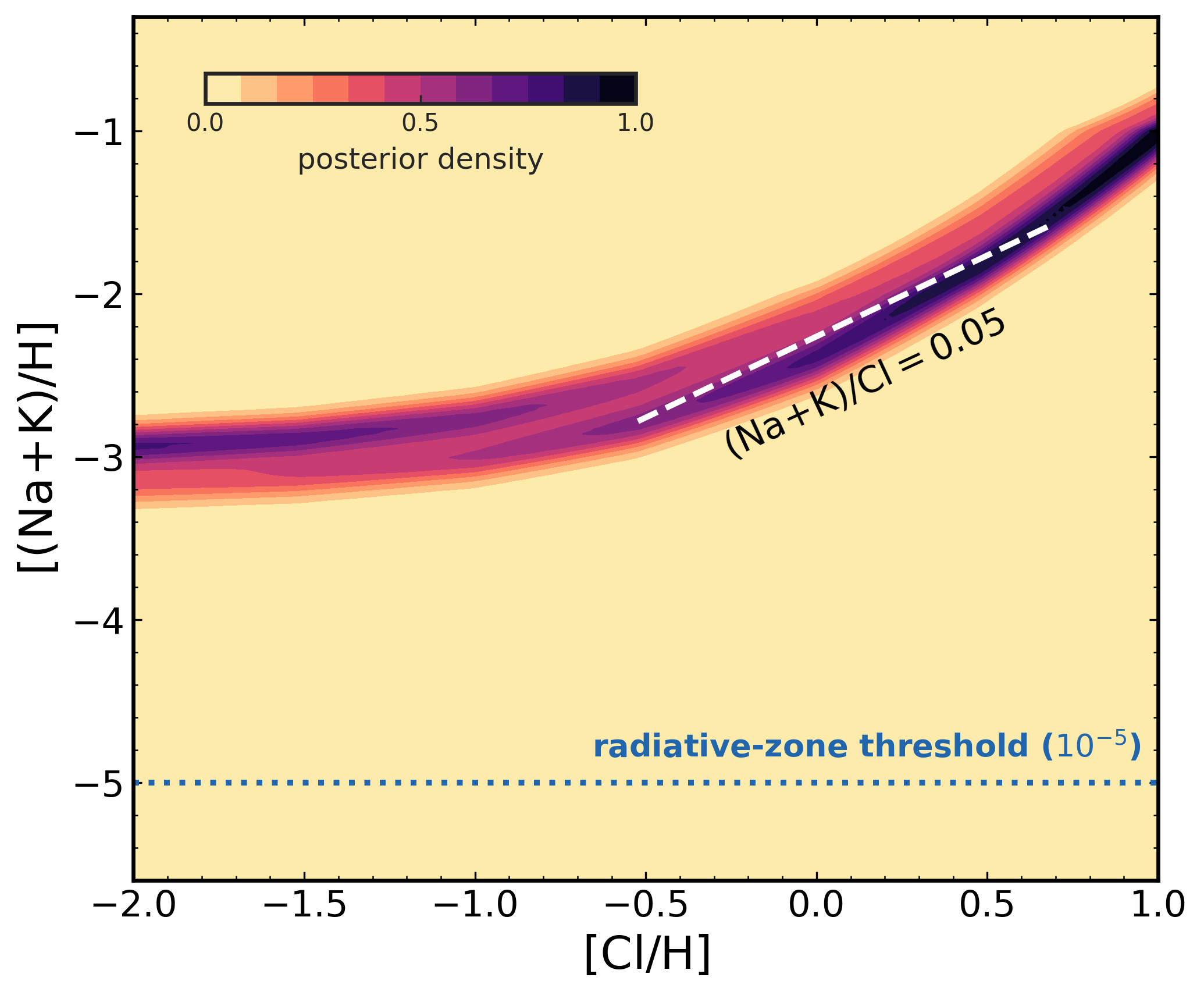}
\caption{Joint posterior distribution of deep gas-phase alkali and chlorine abundances under the
equilibrium-rainout model with sulfur fixed to the \textit{Galileo} value. Axes are the chlorine and
alkali metallicities $[\mathrm{Cl/H}]$ and $[\mathrm{(Na\!+\!K)/H}]$. The white dashed line marks the
best-fit relation, $\nak\approx0.05$
(Equation~\ref{eq:degeneracy}), which holds over $0.3$--$5\xsol$ in chlorine; at lower chlorine the ridge
flattens toward the $\sim1.0\times10^{-3}\xsol$ alkali floor set by the \ce{HS-} sink. The diagonal ridge shows
that the MWR electron density constrains the alkali abundance conditional on chlorine, while neither elemental
abundance is independently measured. The blue dotted line marks the $10^{-5}\xsol$ alkali threshold for a
radiative zone \citep{siebenalerAlkaliLinesExtreme2026}; the posterior ridge lies two to three orders of
magnitude above it across the sampled chlorine abundances. The display priors are flat in $\log\mathrm{alkali}$
and $\log\mathrm{Cl}$.}
\label{fig:result}
\end{figure}

At fixed alkali abundance, increasing chlorine reduces the electron density. Reproducing the observed electron density consequently requires the alkali abundance to increase with chlorine. A linear fit to the equilibrium-chemistry calculation (Figure~\ref{fig:result}) yields
\begin{equation}
\nak\approx0.05,
\label{eq:degeneracy}
\end{equation}
roughly $180$ times below the protosolar ratio of $8.7$, with an alkali floor of $\sim\!1.0\times10^{-3}\xsol$ at low chlorine abundance (below $0.1 \xsol$), set by the residual \ce{HS-} sink rather than by chlorine. Equivalently, in metallicity terms the alkali enrichment lies a factor of $\sim180$ below the chlorine enrichment, $[\mathrm{(Na\!+\!K)/H}]\approx[\mathrm{Cl/H}]-2.3$; the elemental ratio is small because chlorine is much less abundant than the alkalis in the Sun. The relation remains close to proportional for chlorine $0.3$--$5\xsol$; below this range, the alkali abundance approaches the \ce{HS-}-set floor, whereas above it the relation steepens as alkali chlorides begin to condense. This relation is specific to the equilibrium-rainout limit; disequilibrium mineral-cloud processes can alter the charge balance.

The alkali--chlorine degeneracy appears as a diagonal ridge in the joint posterior (Figure~\ref{fig:result}). At fixed chlorine abundance, the well-constrained electron density limits the alkali abundance to a spread of only $\sim\!0.2$~dex. Along the ridge, however, the alkali abundance varies across nearly the full $\sim\!3$~dex chlorine range. For the illustrative case of $3\xsol$ chlorine, comparable to Jupiter's measured volatile enrichment, we infer a deep gas-phase alkali abundance of $1.6\times10^{-2}\xsol$ ($1\sigma$: $1.2\times10^{-2}$--$2.7\times10^{-2}\xsol$), roughly two orders of magnitude below the volatile enrichment. Thus, under equilibrium chemistry, the MWR observations require substantially depleted deep gas-phase alkalis, although the inferred depletion remains conditional on the unknown chlorine abundance through Equation~\ref{eq:degeneracy}.

For the $3\xsol$ reference case, our inferred alkali abundance lies between the two previous estimates derived from the same \textit{Juno} data. It falls at the upper end of the $10^{-2}$--$10^{-5}\xsol$ range inferred by \citet{bhattacharyaHighlyDepletedAlkali2023}, who considered only alkali ionization and electron recombination in a Saha calculation. Under equilibrium chemistry, however, electrons are the residual of the cation--anion balance. Including the dominant anions \ce{Cl-} and \ce{HS-} therefore requires a higher alkali abundance to reproduce the same electron density.

Our estimate is nevertheless about a factor of six lower than the best-fit $\sim\!0.1\xsol$ value reported by \citet{aglyamovAlkaliMetalDepletion2025}. We attribute this difference primarily to the fitting strategy. Their analysis fit the overall $T_b$--$L_d$ correlation, as in \citet{bhattacharyaHighlyDepletedAlkali2023}, assuming horizontally uniform temperature and \ce{NH3}, and adjusted the ammonia opacity and alkali abundance to reproduce the ensemble behavior. In contrast, we fit the full latitude-dependent distributions of $T_b$ and $L_d$ while adopting the latitude-resolved temperature and \ce{NH3} profiles retrieved by \citet{liSuperadiabaticTemperatureGradient2024} above 200~bar. This treatment accounts for the large upper-atmosphere variations and more cleanly isolates the deep-atmosphere signal.
\subsection{Implication for a Deep Radiative Zone}\label{sec:rz}

A stable deep radiative zone requires gas-phase alkali abundances below $\sim\!10^{-5}\xsol$ \citep{siebenalerAlkaliLinesExtreme2026}. No equilibrium solution considered here reaches this threshold. As chlorine is reduced below $0.1 \xsol$, the \ce{HS-} charge sink sets an alkali floor near $10^{-3}\xsol$ while sulfur remains fixed at the \textit{Galileo} value. Even if sulfur and chlorine are both depleted in a uniformly metal-poor envelope, the charge balance approaches the Saha-ionization limit and reproduces the observed electron density at an alkali abundance of $\sim1.3\times10^{-4}\xsol$, still an order of magnitude above the radiative-zone threshold. Our results therefore exclude the previously proposed global radiative zone at kilobar pressures.

\section{Discussion and Conclusions} \label{sec:discussion}

The main result of this work is that the deep electron abundance measured by \textit{Juno} MWR constrains Jupiter's alkali-to-chlorine ratio, rather than its alkali abundance alone. In the equilibrium-rainout limit, the allowed gas-phase abundances follow Equation~\ref{eq:degeneracy}, with an alkali floor set by \ce{HS-} at low chlorine abundance. This result makes chlorine---an element that is inaccessible to direct remote sensing in Jupiter's observable atmosphere---the key compositional uncertainty in the equilibrium interpretation of the MWR data.

Chlorine provides an important but missing link to Jupiter's formation context. The \textit{Galileo} probe found carbon, nitrogen, sulfur, and the noble gases enriched by approximately $3\xsol$ \citep{wongUpdatedGalileoProbe2004,atreyaDeepAtmosphereComposition2020}, establishing that Jupiter's molecular envelope is enriched in volatile material. Chlorine could have accompanied these volatile-rich solids: it can be incorporated into chloride minerals or, after refractory chlorine condenses, as residual \ce{HCl} trapped in water ice \citep{loddersAlkaliElementChemistry1999,loddersSolarSystemAbundances2003,loddersSolarSystemAbundances2023}. Measurements of comet 67P likewise suggest limited chlorine fractionation in Solar System solids \citep{dhoogheHalogensTracersProtosolar2017,altweggEvidenceAmmoniumSalts2020}. However, unlike the \textit{Galileo} volatiles, chlorine cannot be measured in Jupiter's observable atmosphere because \ce{HCl} is removed by \ce{NH4Cl} condensation. Whether chlorine follows the measured volatile enrichments therefore remains unknown.

Previous interpretations of the MWR electron signal inferred depleted alkalis and compared them with the \textit{Galileo} volatile inventory, thereby connecting the electron measurement indirectly to the relative delivery of rock and ice. Our result adds a distinct constraint: under equilibrium chemistry, chlorine is part of the charge balance that sets the electron abundance, so the MWR data constrain the refractory alkalis directly relative to an unmeasured volatile, chlorine. The alkali--chlorine relation thus provides a new refractory-to-volatile diagnostic that does not require assuming that chlorine is enriched by $3\xsol$. A future deep-chlorine constraint would convert this relation into an absolute alkali abundance and allow a more direct test of the composition of Jupiter's accreted solids.

The relation is nevertheless informative when viewed against the \textit{Galileo} inventory. If chlorine is enriched similarly to the measured volatiles, with a chlorine abundance of $\sim3\xsol$, Equation~\ref{eq:degeneracy} implies a gas-phase alkali abundance of only $\sim1.6\times10^{-2}\xsol$. Equivalently, the inferred alkali-to-chlorine ratio, $\nak\approx0.05$, is lower than the protosolar ratio of $8.7$ by a factor of $\sim180$. Such a large refractory-to-volatile contrast would not arise from uniform enrichment by solar-composition solids; it would require chemical fractionation, atmospheric rainout, or a non-equilibrium interpretation of the electron signal. Because the inferred ratio is a gas-phase quantity in one region of an atmosphere with condensation and transport, it should not yet be identified directly with the bulk composition of Jupiter's building blocks. Instead, it exposes a testable tension between the equilibrium interpretation of the MWR data and simple models of uniform solid enrichment.

Furthermore, our interpretation is conditional on the underlying chemistry and atmospheric dynamics. In the disequilibrium mineral-cloud scenario of \citet{zhangAlkaliMetallicity2025}, vertical mixing, mineral chemistry, and dust--plasma interactions can suppress the electron abundance while the bulk alkali abundance remains solar or supersolar. Equation~\ref{eq:degeneracy} does not apply in that limit. Distinguishing whether and how much the mineral clouds diverge from equilibrium rainout is therefore essential before interpreting the MWR electron signal as a constraint on Jupiter's overall refractory-to-volatile composition.

The radiative-zone conclusion is robust to the difference between these interpretations. Under equilibrium rainout, the \ce{HS-} floor prevents the alkali abundance from falling to the $\sim10^{-5}\xsol$ level required for a global radiative zone (Section~\ref{sec:rz}). In the disequilibrium mineral-cloud scenario, the electron abundance can be suppressed even when the bulk alkali abundance is solar or supersolar \citep{zhangAlkaliMetallicity2025}; such an atmosphere has \textit{greater} deep opacity than in the equilibrium case, with additional opacity from the clouds themselves \citep{siebenalerConditionsRadiativeZones2025}. Thus, although bulk alkali abundances can differ greatly between equilibrium rainout and disequilibrium mineral cloud assumptions, neither scenario supports a global kilobar radiative zone. 

Future progress requires tests of both the chlorine abundance and the electron chemistry. Chlorine-bearing species associated with the \ce{NH4Cl} cloud, or transient \ce{HCl} enhancements above vigorous storms, could provide indirect constraints on deep chlorine \citep{showmanHydrogenHalidesJupiter2001,aglyamovAlkaliMetalDepletion2025}. Improved measurements of hot-\ce{NH3} microwave opacity would sharpen the electron constraint, and spatially resolved multichannel MWR analyses coupled to mineral-cloud microphysics could distinguish the equilibrium-rainout and disequilibrium scenarios. These observations could clarify how the MWR electron signal records Jupiter's deep alkali--chlorine chemistry and, ultimately, the balance of rock and ice in the material that formed Jupiter.

\begin{acknowledgments}
We thank Fabiano Oyafuso and Zhimeng Zhang for providing the MWR data. X.Z. is supported by the
National Science Foundation Astronomy and Astrophysics Research grant (AAG) 2307463, the NASA
Exoplanet Research grant (XRP) 80NSSC22K0236, and the NASA Interdisciplinary Consortia for
Astrobiology Research grant (ICAR) 80NSSC21K0597. C.L. and J.H. are supported by NASA's
\textit{Juno} project NNM06AA75C and a subaward to the University of Michigan with project
No.\ Q99063JAR.  Y.A. is supported by NASA Research grant 80NSSC26K0129. L.S is supported by the project ENW.GO.001.001 of the research programme “Use
of space infrastructure for Earth observation and planetary research (GO), 2022-
1” which is (partly) financed by the Dutch Research Council (NWO). LLM tools (Opus 4.8) were used to assist with coding and writing. 
\end{acknowledgments}

\software{GGchem \citep{woitkeEquilibriumChemistry1002018},
HARP \citep{liHighperformanceAtmosphericRadiation2018}.}


\end{document}